\documentclass[fleqn,usenatbib]{mnras}

\usepackage{newtxtext,newtxmath}

\usepackage[T1]{fontenc}

\DeclareRobustCommand{\VAN}[3]{#2}
\let\VANthebibliography\thebibliography
\def\thebibliography{\DeclareRobustCommand{\VAN}[3]{##3}\VANthebibliography}

\usepackage{graphicx}	
\usepackage{amsmath}	

\title[Dynamical simulations in multiple-planet systems]{Do instabilities in high-multiplicity systems explain the existence of close-in white dwarf planets?}

\author[R. F. Maldonado et al.]{
R. F. Maldonado,$^{1}$, E. Villaver $^{2,3}$, A. J. Mustill $^{4}$, M. Ch\'avez $^{1}$, E. Bertone $^{1}$  \\
\thanks{E-mail: raulfms@inaoep.mx}
$^{1}$Instituto Nacional de Astrof\'isica, \'Optica y Electr\'onica, Luis Enrique Erro 1, Tonantzintla, 72849, Puebla, M\'exico\\
$^{2}$Departamento de F\'isica Te\'orica, Universidad Aut\'onoma de Madrid, 28049 Madrid, Spain\\
$^{3}$Centro de Astrobiolog\'ia (CAB, CSIC-INTA), ESAC Campus Camino Bajo del Castillo, s/n, Villanueva de la Ca\~nada, 28692, Madrid, Spain\\
$^{4}$Lund Observatory, Box 43, SE-22100 Lund, Sweden\\
}

\date{Accepted XXX. Received YYY; in original form ZZZ}

\pubyear{2020}

\begin{document}
\label{firstpage}
\pagerange{\pageref{firstpage}--\pageref{lastpage}}
\maketitle

\begin{abstract}
We investigate the origin of close-in planets and related phenomena orbiting white dwarfs (WDs), which are thought to originate from orbits more distant from the star. 
We use the planetary architectures of the 75 multiple-planet systems (four, five and six planets) detected orbiting main-sequence stars to build 750 dynamically analogous templates that we evolve to the WD phase.  Our exploration of parameter space, although not exhaustive, is guided and restricted by observations and we find that the higher the multiplicity of the planetary system, the more likely it is to have a dynamical instability (losing planets, orbit crossing and scattering), that eventually will send a planet (or small object) through a close periastron passage. Indeed, the fraction of unstable four- to six-planet simulations is comparable to the 25--50$\%$ fraction of WDs having atmospheric pollution. Additionally, the onset of instability in the four- to six-planet configurations peaks in the first Gyr of the WD cooling time, decreasing thereafter. Planetary multiplicity is a natural condition to explain the presence of close-in planets to WDs, without having to invoke the specific architectures of the system or their migration through the von Zeipel--Lidov--Kozai (ZLK) effects from binary companions or their survival through the common envelope phase. 
\end{abstract}




\section{Introduction}

Planets commonly found orbiting main sequence (MS) stars are in jeopardy as their host stars evolve off the MS to the red giant and asymptotic giant branch phases \citep{villaver2007}. Planet engulfment during the giant phases of the star is guaranteed to occur to certain orbital distances set by a combination of tidal interaction and stellar mass-loss  \citep{villaver2009,kunitomo2011,mustill2012,nordhaus2013,villaver2014}. This has a strong dependency on the stellar mass and is quite sensitive to the planetary mass  \citep[e.g.,][]{villaver2014}. Several lines of evidence show, however, that both planets and rocky bodies must be dynamically delivered to the white dwarf's (WD) proximity after the end of the asymptotic giant branch (AGB). The first direct transit detection of a Jovian-size planet candidate in a compact 1.4-day orbit has just been found \citep{vanderburg2020}, and related
observed phenomena such as atmospheric pollution \citep[e.g.,][]{zuckerman2003}, near-infrared excesses \citep[e.g.,][]{gansicke2006}, asteroids \citep{vanderburg2015,manser2019,vanderbosch2020}, or even planets inferred from a gas disc \citep{gansicke2019} have been reported for years.  

As a star becomes a WD, it loses a considerable fraction of its mass; this moves the planets onto wider orbits, but also leads to an increase of the planet:star mass ratio, which radically changes the dynamics and stability of multiple planetary systems. The pioneering work of \citet{debes2002} that explored the effects of mass-loss on dynamical evolution has been followed by many studies of two-planet systems \citep{voyatzis2013,veras2013, veras2013b,mustill2014,smallwood2018,ronco2020,maldonado2020}. However, similar studies using three and four planets in generic planetary systems are more scarce \citep{veras2015,veras2016b,mustill2018,maldonado2020b}.  The overall conclusion of all these works is that instabilities that lead to the loss of a planet in two- and three-planet systems do not occur often enough to explain the high incidence of atmospheric pollution observed in WDs, even when considering instabilities such as orbit crossing and orbital scattering. 

The point we make in the present Letter is that a high multiplicity of planets increases the planet--planet scattering events that lead to dynamical instabilities that eventually pollute WDs and drive planets and/or asteroids to its close vicinity.  Furthermore, we provide a natural mechanism to explain the presence of planets reaching close orbits round WDs, without having to invoke the specific architecture of the system as a hierarchical quadruple to explain its migration through the von Zeipel--Lidov--Kozai (ZLK) effect \citep{zeipel1910,lidov1962,kozai1962}, as has been done for the recently discovered WD~1856~b \citep{vanderburg2020} in \citet{connor2020,munoz2020,alexander2020}. 
 For the first time, we study the post-MS dynamical evolution of four- to six-planet systems by using scaled versions of the planetary architectures observed orbiting MS stars. We also expand the previously explored parameter space.

\section{Numerical simulations set-up}

In \citet[][Papers~I and~II from now onwards]{maldonado2020,maldonado2020b}, we evolved the planetary architectures of the two- and three-planet systems. Here, we follow the same procedure to simulate the evolution of higher multiplicity (four, five, and six) detected planetary systems orbiting MS stars. To do so, first, we 
obtain from the Exoplanet Encyclopedia \citep{schneider2011},\footnote{http://exoplanet.eu/} and the NASA Exoplanet Archive \citep{akeson2013} \footnote{https://exoplanetarchive.ipac.caltech.edu/} 51 four-planet systems, 14 five-planet systems, and 10 six-planet systems with reported discovery up to September 2020, excluding from this list close binaries and pre-MS stars. The final sample used in this work contains a total of 75 MS stars. From the observations,  we select the planet and stellar masses, radius, and eccentricity, when available. If not available, as in Papers\, I and \,II, we have used a standard planet mass-radius relation \citep{chen2017},  and Rayleigh distributions with a $\sigma$ parameter $\sigma=0.02$ \citep{pu2015} for the eccentricity, and with $\sigma = 1.12^\circ$ \citep{xie2016} for the inclination angles. 
For each observed template system, we run 10 clones, differing in the randomisation of their orbital phases and orientations.

The host stars that originally span a mass range from $0.27$ to $1.56\mathrm{\,M}_\odot$ have been re-scaled  to a 3 $\mathrm{M}_\odot$ mass star, while keeping them dynamically analogous to the original system.  A 3 $\mathrm{M}_\odot$ star represents the average MS mass of observed polluted WDs ($\sim$0.7 $\mathrm{M}_\odot$, \citealt{koester2014}) and allows us to follow the evolution to the WD in a feasible computational time. If lower-mass progenitors were considered, we would expect slightly fewer systems to become unstable, as they lose a slightly smaller fraction of their mass in becoming a WD \citep{mustill2014}. The evolution of the star has been computed using the SSE code \citep{hurley2000} assuming the Reimers mass-loss parameter $\eta=0.5$ and solar metallicity. The planetary masses have been re-scaled as well, to conserve the dynamical properties (multiplying the planet mass by the ratio between the modelled and the observed stellar mass in the system).

We use the \textsc{Mercury} package \citep{chambers1999} modified by \citep{veras2013b,mustill2018} with an implemented RADAU integrator with a tolerance parameter of 10$^{-11}$ as in \cite{mustill2018} and Papers \,I, \,II  to evolve the planetary systems from the start of the MS for 10 Gyr. Planets are removed from the simulations when they reach an orbital distance of $1\times10^6$ au from the central star, which we consider an ejection. Planets colliding with each other or with the stellar radius are also removed. To avoid the complications of tidal evolution on the AGB and ensure their survival to engulfment, we placed the innermost planet (planet\,1) at a semimajor axis $a_0$ = 10\,au (see \citealt{villaver2009,mustill2012}). We assure the conservation of the dynamics of the system by placing consecutive pairs at distances  $(a_j/a_1)a_0$ (where $a_1$, and $a_j$ are the observed semimajor axes of planets\,1 and $j$ (with $2 \leq j \leq 6$) respectively.

In Fig. \ref{mass-ecc} (see appendix), we present histograms of the planet mass distribution (left-hand panel) and the eccentricity distribution (middle panel) of the simulated planets using different colours according to the number of planets in the system. We have added in (light) black, for reference, the distribution of the two- and three-planet systems simulated in Papers\,I and \,II.   The right-hand panel of Fig. \ref{mass-ecc} displays the distribution of $\Delta$ (in mutual Hill units defined as $\Delta=(a_j-a_i)/R_\mathrm{m,Hill}$, where i,j indicates a planet pair i-j and $R_\mathrm{m,Hill}$ is the mutual Hill radius of the planet pair) for adjacent pairs of planets.

\section{Results}

We have performed a total of 750 simulations (10 simulations per system configuration) where 510, 140 and 100 are of four-, five- and six-planet systems respectively. We have removed from the following analysis all the planetary systems that have at least one pair of planets in mean motion commensurabilities, which produce planet losses and orbit crossing on the MS.  That is the case for a total of 167 simulations (105 are in 12 systems with four planets, 12 in 3 systems with five planets, and 50 in 6 systems of six planets).

\begin{figure}
\begin{center}
\begin{tabular}{cc}
\includegraphics[width=8cm, height=6cm]{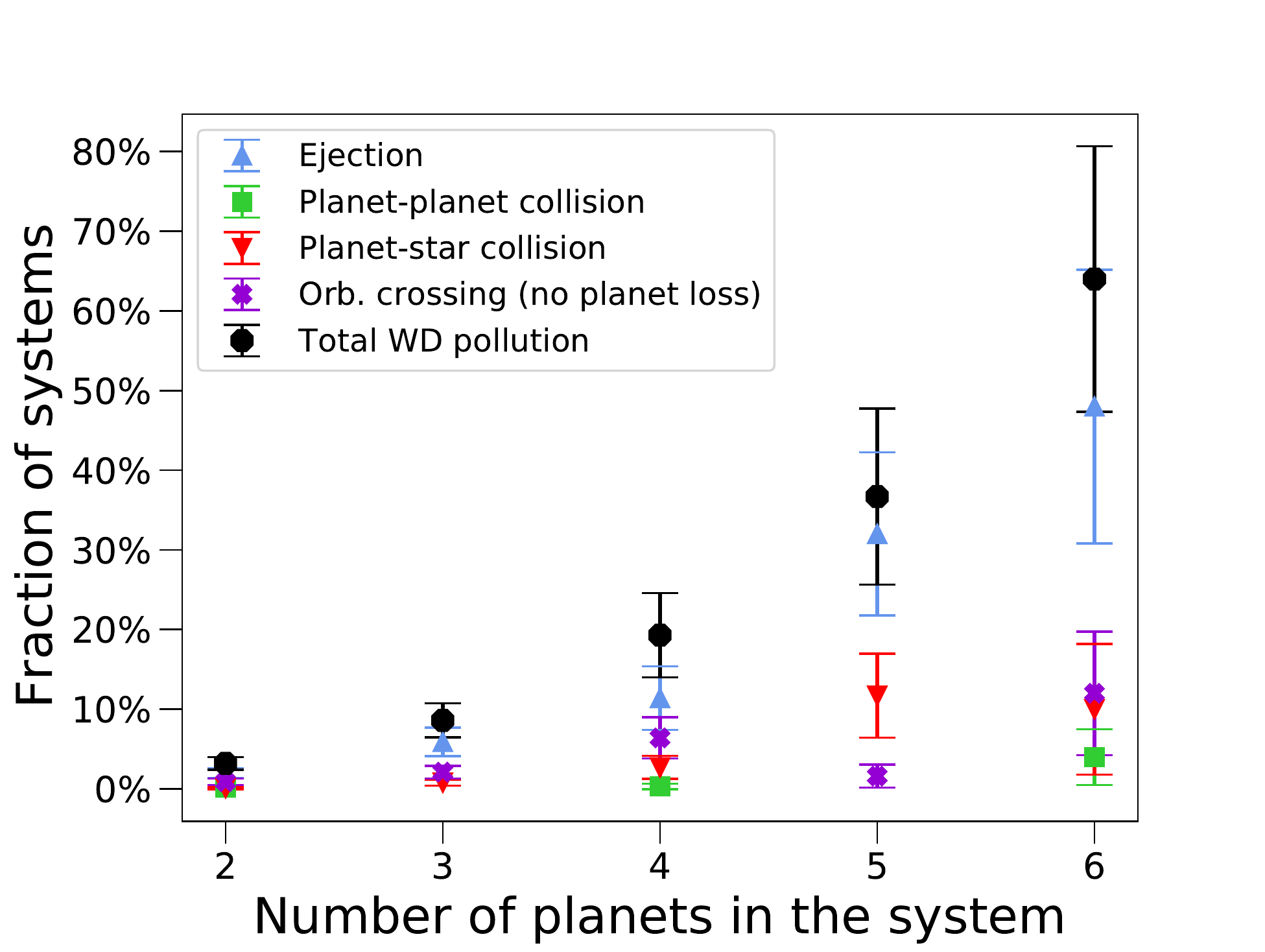}
\end{tabular}
\caption{Relative number of multiple-planet systems with respect to the total number of simulations losing planets by Hill and/or Lagrange instabilities as well as the simulations with orbit crossing that do not lose any planet during 10 Gyr. Additionally, we present the total number of simulations that may contribute to WD pollution. Error bars correspond to the standard deviation of the Bootstrap resampling method.  Each instability type is marked with a different symbol and colour. }
\label{insfig}
\end{center} 
\end{figure}

In Table \ref{tabwd}, we present the numbers and percentages of planets lost by different dynamical instabilities on the WD phase. We count only different simulations where at least one planet is lost. In Fig. \ref{insfig}, we display the fraction of systems losing planets either by Hill (planets collide with each other or their orbits cross) or Lagrange instabilities (ejected or collide with the star) with respect to the number of simulations in each planetary configuration (including those from Papers \,I, \,II). We see that as the multiplicity of planets increase, the fraction of instabilities also increases. 

It is important to note that planet ejections not only happen more frequently in our four- to six-planet simulations, but also that the surviving planets remain dynamically active after the ejection, having eccentricity excitation that produces several orbit crossings.  A planetesimal belt, if present, could be disrupted that way. We have a non-negligible number of simulations where, despite no planets being lost, there is orbit crossing and orbital scattering during the WD phase. We have listed those in Table \ref{tabwd} and added them to the total contribution to the WD pollution in the last line of the table. Indeed \cite{mustill2018} show that planets having a mass $\leq$ 100 M$_\oplus$ (low-mass planets) are efficient deliverers of planetesimals toward the WD. The percentage of our simulations with low-mass planets involved in simulations dynamically active is high: $\sim$ 65\,$\%$ for planet masses in the range of 0.9--85 M$_\oplus$ in the four- and five-planet system simulations and low, $\sim$10\,$\%$ in the six-planet system case.

\begin{figure}
\begin{center}
\begin{tabular}{cc}
\includegraphics[width=8.5cm, height=6.7cm]{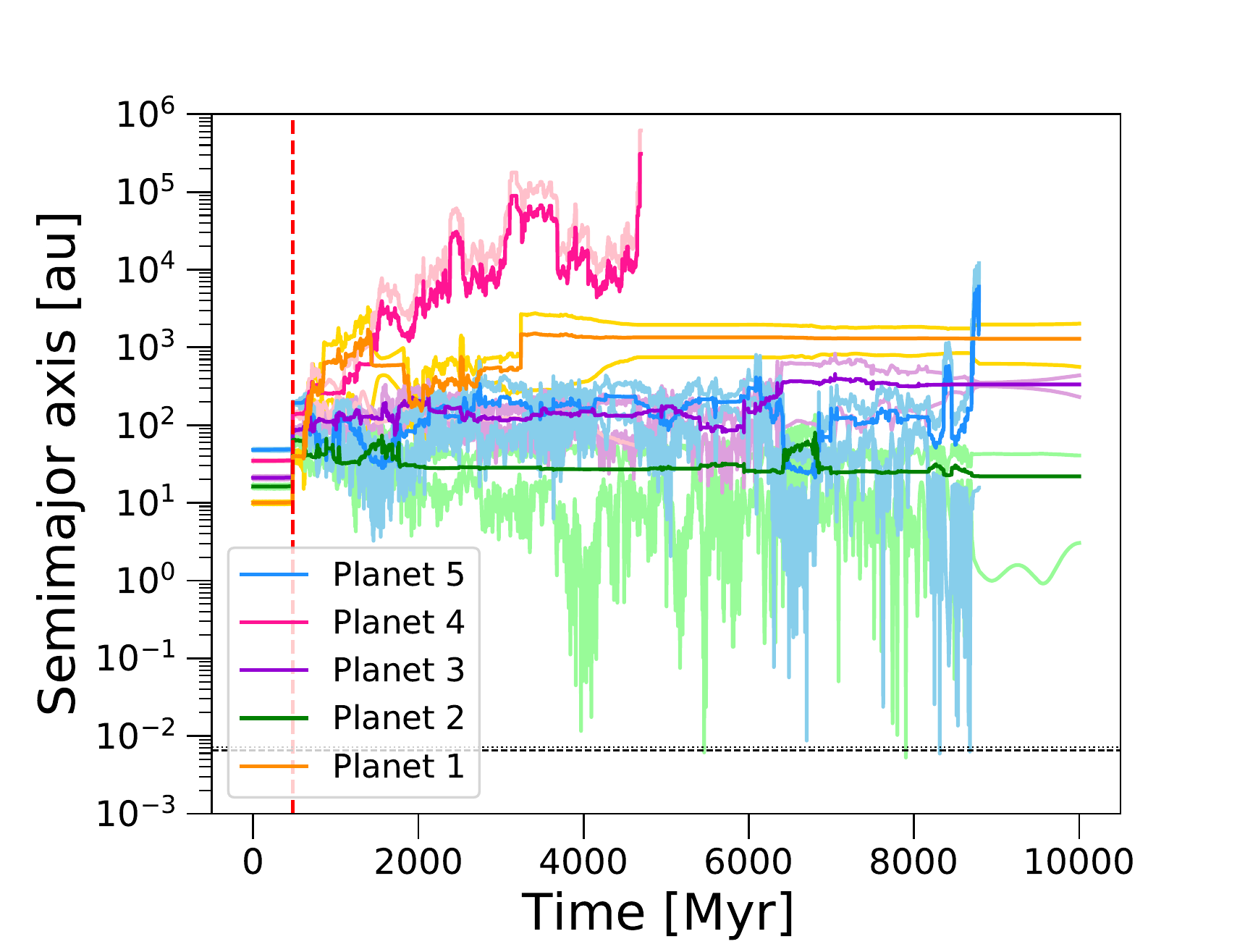}
\end{tabular}
\caption{Semimajor axis evolution of the scaled five-planet system {\it Kepler}-84. Planets are displayed with different colours, with the lighter version of the colors showing the pericentre and apocentre. The red vertical dashed line indicates the beginning of the WD phase. The black dotted and dashed lines show the WD Roche radius for planet\,2 (mass 21.7 M$_\oplus$) and 5 (mass 68.6~M$_\oplus$), respectively. }  
\label{exam}
\end{center} 
\end{figure}

Planets can be sent to very close distances (even crossing the Roche radius) the of the WD. As an example of a system with a complex dynamical behaviour, we show in Fig. \ref{exam} the scaled system {\it Kepler}-84: several orbit crossings occur among the five planets, producing the ejections of planets~4 and~5, while planets~2  and~5  have such a high eccentricity excitation that cause both planets to have several pericentre passages within the WD Roche radius.  The number of simulations that have at least one planet having pericentre passages up to 10 Roche radii of the WD is listed in Table \ref{tabwd} as well as the number of them that indeed cross the Roche radius. Some simulations have more than one planet crossing the Roche radius. Reported in the planet--star collision row of Table \ref{tabwd} are the planets that entered the Roche radius and finally collided with the WD.

\begin{table*} 
\small\addtolength{\tabcolsep}{-3pt}
\begin{center}
\caption{Instabilities at the WD phase according to the number of planets in the simulated system. The first percentage is given with respect to the total number of simulations and the second one is the fraction with respect to the total number of planets. Note that several planets are lost in the same simulation and some with different instability type, however the total number of systems reflects only different simulations where a planet is lost.}
\label{tabwd}
\resizebox{0.97\textwidth}{!}{%
\begin{tabular}{l r r r r r r r r r r}
\noalign{\smallskip} \hline \noalign{\smallskip}
& \multicolumn{2}{c}{Two-planet systems} & \multicolumn{2}{c}{Three-planet systems} & \multicolumn{2}{c}{Four-planet systems} & \multicolumn{2}{c}{Five-planet systems} & \multicolumn{2}{c}{Six-planet systems}\\
 & Systems & Planets & Systems & Planets & Systems & Planets & Systems & Planets & Systems & Planets \\
\noalign{\smallskip} \hline \noalign{\smallskip}
{\bf Ejections}  & 68 (2.0\,$\%$) & 68 (1.0\,$\%$) & 69 (5.9\,$\%$) &  71 (2.0\,$\%$) & 46 (11.4\,$\%$)& 67 (4.1\,$\%$) & 41 (32.0\,$\%$) & 74 (11.6\,$\%$) & 24 (48.0\,$\%$) & 57 (19.0\,$\%$) \\
{\bf Planet--star collisions} & 5 (0.1\,$\%$) & 5 (0.1\,$\%$) &  9 (0.8\,$\%$) & 9 (0.3\,$\%$) & 11 (2.7\,$\%$) & 11 (0.7\,$\%$) & 15 (11.7\,$\%$) & 15 (2.3\,$\%$) & 5 (10.0\,$\%$) & 5 (1.7\,$\%$) \\
{\bf Planet--planet collisions} & 7 (0.2\,$\%$) & 7 (0.1\,$\%$) &  -- &  -- & 1 (0.3\,$\%$)   & 1 (0.1\,$\%$) & -- & -- & 2 (4.0\,$\%$) & 2 (0.7\,$\%$)\\
{\bf Planets within 10 Roche radii} &  8 (0.2\,$\%$) & 8 (0.1\,$\%$) &  26 (2.2\,$\%$) & 30 (0.9\,$\%$) & 29 (7.2\,$\%$)  & 47 (2.9\,$\%$) & 32 (25.0\,$\%$) & 48 (7.5\,$\%$)  & 14 (28.0\,$\%$) & 23 (7.7\,$\%$)\\
{\bf Planets within 1.4 day orbits} & 8 (0.2\,$\%$) & 8 (0.1\,$\%$) &  22 (1.9\,$\%$) &  26 (0.8\,$\%$) & 27 (6.7\,$\%$)  &  40 (2.5\,$\%$) & 32 (25.0\,$\%$) & 47 (7.3\,$\%$)  & 12 (24.0\,$\%$) & 18 (6.0\,$\%$)\\
{\bf Planets entering the Roche radius} & 7 (0.2\,$\%$) & 7 (0.1\,$\%$) &  19 (1.6\,$\%$) &  22 (0.6\,$\%$) & 23 (5.7\,$\%$)  &  33  2.0\,$\%$) & 30 (23.4\,$\%$) & 45 (7.0\,$\%$)  & 12 (24.0\,$\%$) & 16 (5.3\,$\%$)\\
{\bf Orbit crossing and planet losses} & 26 (0.8\,$\%$) & -- &  63 (5.4\,$\%$) &  -- & 52 (12.8\,$\%$)   & -- & 45 (35.2\,$\%$) & -- & 26 (52.0\,$\%$) & --\\
{\bf Orbit crossing and/or scattering} & 32 (0.9\,$\%$) & -- &  24 (2.1\,$\%$) &  -- & 26 (6.4\,$\%$)   & -- & 2 (1.6\,$\%$) & -- & 6 (12.0\,$\%$) & --\\
{\bf without planet losses} &  &  &  &  &  &  &  &  &  & \\
{\bf Total (losing planets)} & 80 (2.3\,$\%$) & 80 (1.2\,$\%$) & 76 (6.6\,$\%$) & 80 (2.3\,$\%$) & 52 (12.8\,$\%$)& 79 (4.9\,$\%$) & 45 (35.2\,$\%$) & 89 (13.9\,$\%$) & 26 (52.0\,$\%$) & 64 (21.3\,$\%$) \\
{\bf Total (WD pollution)} & 112 (3.2\,$\%$) &  & 100 (8.6\,$\%$) &  & 78 (19.3\,$\%$) &   & 47 (36.7\,$\%$) &  & 32 (64.0\,$\%$) &  \\
{\bf Total (for statistics)} & 3485  & 6970 & 1160  & 3480 & 405 & 1620  & 128 & 640 & 50 & 300 \\
\hline
\end{tabular}}
\end{center}
\end{table*}



\begin{figure*}
\begin{center}
\begin{tabular}{cc}
\includegraphics[width=15cm, height=6.0cm]{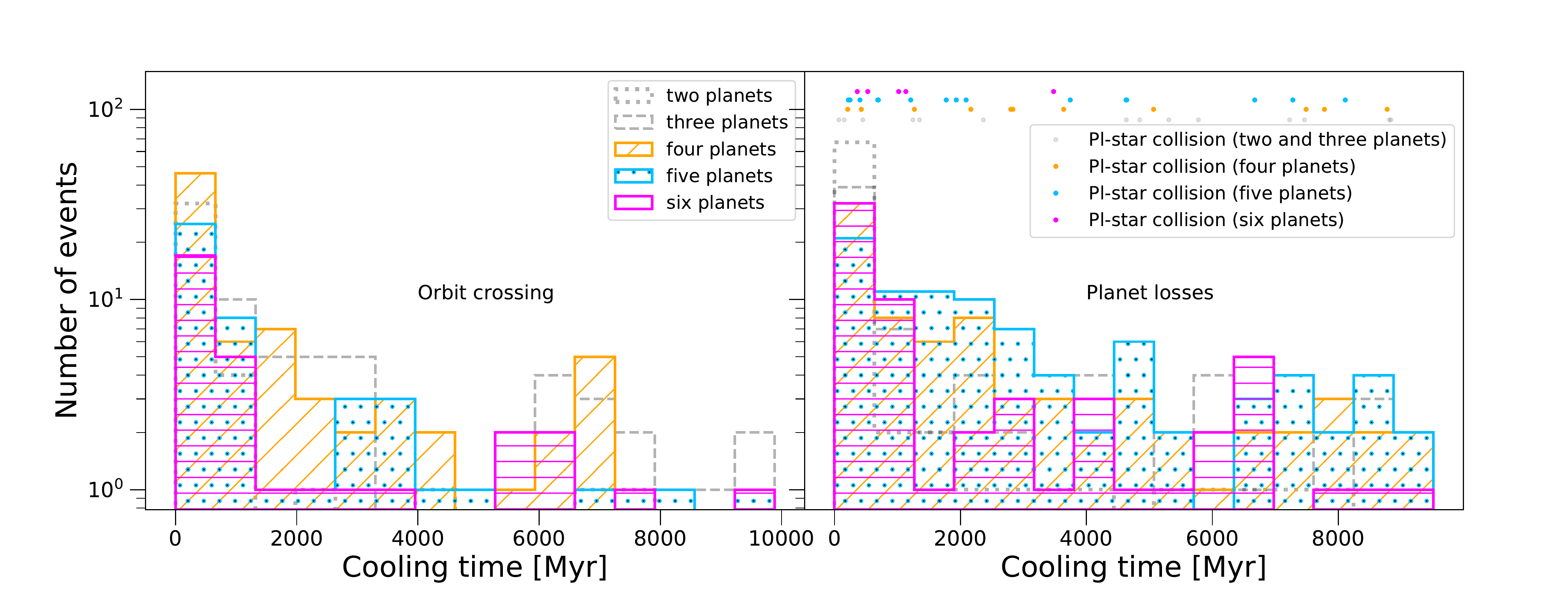}
\end{tabular}
\caption{Left-hand panel: Distribution of cooling time of the first orbit crossing. Right-hand panel: The same as the left-hand panel but showing the cooling time when planets are lost by Hill or Lagrange instabilities. The dots in the upper part indicate the time when planet-star collisions happen and different colours refer to systems with different number of planets. }
\label{coolins}
\end{center} 
\end{figure*}

The times at which the instability events happen are shown in Fig. \ref{coolins}. The left-hand panel is the cooling time of the first orbit crossing with and without planet losses, and the right-hand one reflects when the planet losses happen. We have a peak in the number of orbit crossing and planet losses in the first Gyr but with the orbit crossing events slowing down as the WD ages (only an slight increase around 7 Gyr in the four-planet case) and the planet losses maintained at lower rates throughout the simulated times. Most of the planet losses are ejections but, as mentioned before, the surviving planets continue being dynamically active.  In the upper part of the right-hand panel, we show the cooling time when planet-star collisions happen. We see that this outcome spans up to 8 Gyr of cooling time (four- and five-planet systems) and only systems with six planets have collisions restricted to 4 Gyr. The time of the destabilizing events has a dependency on the mass of the planets involved. Multi-planet systems having at least one planet with M $>$ 100 M$_\oplus$ tend to destabilize earlier than their low-mass counterparts. As an example the four-planet case, orbit crossing events have a median time of 53.9 Myr  and 1.08 Gyr in the high-mass and low-mass case, respectively (same numbers for the planet losses are 454.5 Myr and 1.54 Gyr, respectively).

\section{Discussion}

Previous studies on the post-MS evolution of planetary systems of high multiplicity are very limited and mostly restricted to four-planet systems. For instance, \citet{veras2015,veras2016b} obtained a much larger fraction of unstable (here unstable includes losing planets and/or having orbit crossing on the WD phase and excludes the MS unstable systems) four-planet system simulations (44.7\,$\%$ and 73.0\,$\%$, respectively) than this work (19.3\,$\%$). Furthermore, all the 19 simulations of six and eight-planets in \citet{veras2016b}  and the 40 simulations that \citet{veras2015} performed using 10 Earth mass planets become unstable in the WD phase. The main reason behind the different percentages obtained is easily appreciated in the right panel of Fig. \ref{mass-ecc}. Previous works have probed packed systems (maximum 15 in mutual Hill radii for terrestrial planets and 8 for Jovian planets) whereas we have used adjacent planet pairs with $\Delta$ values up to 66.3 mutual Hill radii. Note that only 36\,$\%$ of our tested four-planet pairs are in the range between 4-15 mutual Hill radii tested previously and 12\,$\%$ have planets in adjacent pairs with $\Delta$ between 6 and 10 in the six-planet systems. 

The majority of planets in the WD phase are lost  due to ejections in our simulations (see Table \ref{tabwd}) and that is what has been found as well in previous works where planet ejections are the most prominent instability outcome \citep{veras2013,veras2013b,mustill2014,veras2016b,mustill2018,maldonado2020,maldonado2020b}.  The percentage of unstable simulations (where planets are lost or are kept the entire simulated time but do experience orbit crossing and/or orbital scattering) increases with the multiplicity of the planetary system. We cannot attribute this result to having simulated more packed systems (see Fig. \ref{mass-ecc}) nor to small number statistics. We see a significant trend in systems experiencing instability with multiplicity (see Figure~\ref{insfig})\footnote{We determine the errors on the instability fractions by means of bootstrap resampling. Because different simulations based on the same template system cannot be regarded as completely independent, we perform a ``two-layer'' bootstrap, first drawing with replacement from the template systems, and then drawing with replacement from the simulation results for each resampled system. This leads to somewhat larger errors than are obtained by pooling all the simulation results and treating them independently}. Our results show that a higher multiplicity of planets may seem to increase the chance of WD pollution, as the destabilised planets can destabilize asteroid belts, sending the asteroids toward the WD and polluting its atmosphere. 

Recently, the first transiting planet has been discovered in a very tight orbit around a WD \citep{vanderburg2020}: a candidate orbiting the star WD~1856+534 with a mass $\leq$ 11.7 $M_J$ and with a period of 1.4 days. The evolution of the star prevents this planet from being there in such a close orbit \citep[see e.g.][]{villaver2007,villaver2014,mustill2012}, unless it survives a common envelope phase \citep{lagos2020}. Thus, dynamical instabilities among outer bodies must be invoked to explain its presence. 

We find in our simulations several planets that approach the surroundings of the WD following dynamical instabilities (note the closest planet is always located at 10\,au at the start of the simulations). The number of planets reaching pericentre distances $\leq$ 10 Roche radii (listed in Table \ref{tabwd}) is large: 46 planets (with a mass range of 3.3--35.2 M$_\oplus$) and one planet with a mass of 460.8 M$_\oplus$ in the four-planet systems; 48 planets with masses between 1.5 and 68.6 M$_\oplus$ in the five-planet systems, and 23 planets covering a mass range of 6.2--106 M$_\oplus$ in the six-planet systems simulations. 

Only one massive planet M > 1 $\mathrm{M}_J$ crosses the 10 Roche radii threshold despite constituting 6.4$\%$ of the simulated planets. If we restrict the masses to $\geq0.1\mathrm{\,M_J}$ that represent 33.9\,$\%$ of the total planets simulated, we find 24 planets (0.9\,$\%$ of the sample) that reach 10 Roche radii. We find that mostly super-Earth to Saturn-like planets reach close WD distances.  Finally, 7.2, 25\,$\%$ and 28\,$\%$ of simulations have pericentre passages $\leq$ 10 Roche radii and   5.7\,$\%$, 23.4\,$\%$ and 24\,$\%$ do cross the Roche radius in the four-, five- and six-planet systems, respectively. We have obtained a higher but comparable fraction of simulations with planets crossing the Roche radius in the four-planet systems with respect to the two- and three-planet case (less than 2\,$\%$ found in Paper\,I and II). However, the fraction is higher ($\sim$ 24\,$\%$) in the five- and six-planet case.

Our exploration of parameter space, although not exhaustive, is guided and restricted by observations. We find that the higher the multiplicity of the planetary system, the more likely it is to have a dynamical instability that eventually will send a planet (or small object) onto an orbit with a small periastron distance. A similar conclusion was reached by \citet{veras2015} in a much more limited analysis. We also find that small planets are more likely to be destabilized onto an orbit passing close to the WD and therefore most likely to be present and/or disrupted and destroyed by tidal forces, leading to the formation of debris or gaseous disks that could eventually pollute the WD atmosphere \citep{manser2019, veras2019}. 

Note that several studies have attempted to predict the number of planetary systems harboring multiple planets since observations are biased. \citet{zink2019} provide corrections to the incomplete transit multiplicity of the \textit{Kepler} data and estimate that the average number of planets per GK dwarf within the radius and period parameter space of Kepler to be $5.86\pm0.18$. Also 32.3$\pm$2.7$\%$ of solar-like stars should contain at least 8 planets within 500 days.

\section{Conclusions}

We have performed 750 dynamical simulations of 75 detected MS planetary systems with four, five and six planets (10 simulations per system configuration). We find that when increasing the multiplicity of the planetary systems studied, the planet-planet scattering events (losing planets, orbit crossing and scattering) that may contribute to WD pollution (enhanced if putative asteroid belts are present within the system) increase as well.  Indeed, the fraction of four- to six-planet unstable simulations is comparable to the 25--50$\%$ fraction of WDs having atmospheric pollution \citep{zuckerman2003,koester2014,wilson2019}. Additionally, the orbit crossing and planet losses in the four-to-six planetary configurations peak in the first Gyr of the WD cooling time, decreasing as the WD ages.   

Our multiple-planet simulations resulted in a non-negligible fraction of planets reaching orbits within the 10 Roche radii of the WD (some of them crossing the Roche radius and even colliding with the WD), especially in systems with five and six planets with planet masses in the range between super-Earth and Saturn. This confirms that planet-planet scattering in multiple-planet systems leads to the existence of close-in planets around the WD, possibly explaining the origin of the newly discovered WD~1856~b orbiting in a 1.4 days period \citep{vanderburg2020}. Furthermore, these planets or asteroids can be influenced by tidal forces of the WD \citep{veras2019,veras2019b} forming debris disks \citep{gansicke2006,Kilic2007,farihi2016,wilson2019}, with some of them having gaseous components \citep{melis2012,manser2016,manser2020,melis2020}; producing photo-evaporation of their atmospheres \citep{gansicke2019} or having the rocky bodies themselves \citep{vanderburg2015,manser2019,vanderbosch2020} producing WD atmospheric pollution. 

\section*{Acknowledgements}

This research has made use of the NASA Exoplanet Archive, which is operated by the California Institute of Technology, under contract with the National Aeronautics and Space Administration under the Exoplanet Exploration Program.  This research has made use of the SIMBAD database, operated at CDS, Strasbourg, France.  E.V. and R.M. acknowledge support from the `On the rocks II project' funded by the Spanish Ministerio de Ciencia, Innovaci\'on y Universidades under grant PGC2018-101950-B-I00 and the Unidad de Excelencia ``Mar\'ia de Maeztu''- Centro de Astrobiolog\'ia (CSIC/INTA). MC, RM and EB thank CONACyT for financial support through grant CB-2015-256961. A.J.M. acknowledges support from the starting grant 2017-04945 `A unified picture of white dwarf planetary systems' from the Swedish Research Council. We are grateful to Francisco Prada for the use of the computer cluster, to Paul McMillan for statistical advice and to the referee for giving helpful comments to improve the manuscript.

\section*{Data Availability}

The data underlying this article will be shared on reasonable request to the corresponding author.



\bibliographystyle{mnras}
\bibliography{bib} 




\appendix

\section{Additional material}

\begin{figure*}
\begin{center}
\begin{tabular}{cc}
\includegraphics[width=18.5cm, height=6.7cm]{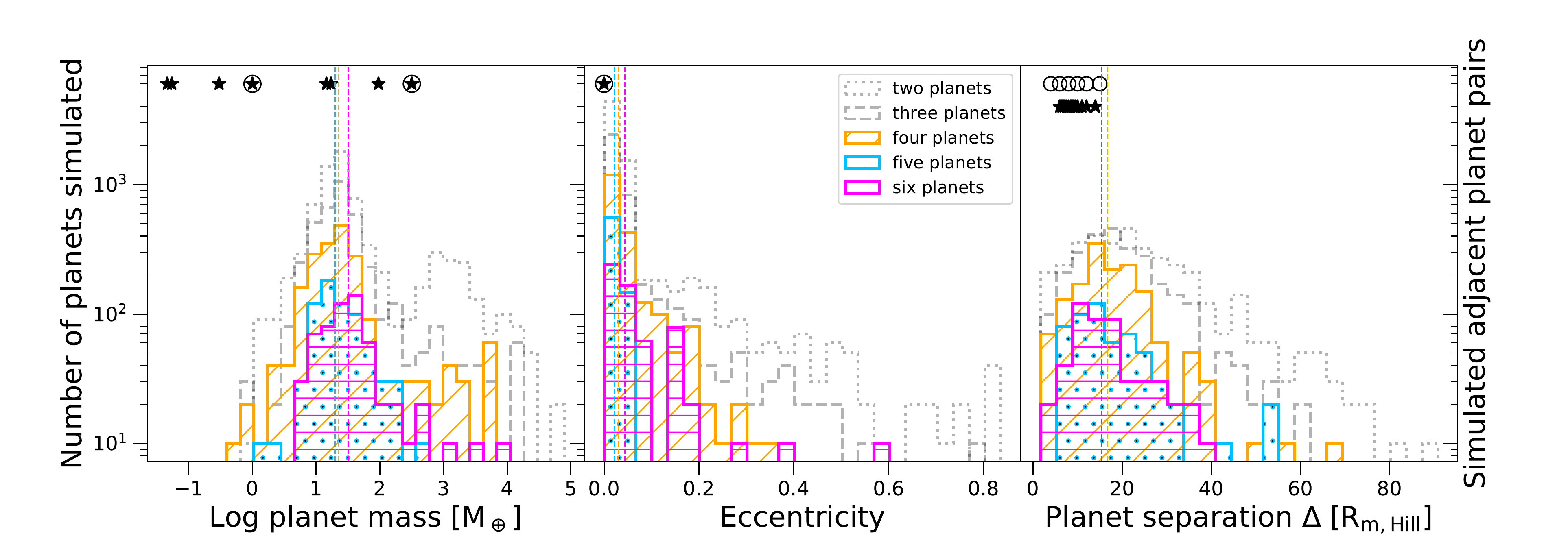}
\end{tabular}
\caption{Left-hand panel: Histograms of the simulated planet masses, different colours are for the planetary system according to the number of planets (see legend). Middle panel: same as the left-hand panel but for the planet eccentricities. Right-hand panel: planet separation $\Delta$ in mutual Hill units of adjacent pairs of planets.  Symbols in the upper part are the values used in previous works: black starry and circles for \citet{veras2016b} and \citet{veras2015}, respectively. The medians of the planet mass (plotted as dashed vertical lines) are 22.6, 19.7 and 32.0 M$_\oplus$ for the four-, five- and six-planet systems, respectively. For the eccentricity: four-planet, five-planet and six-planet systems have a median of 0.03, 0.02 and 0.05, respectively.  The medians of the planet separation $\Delta$ of the adjacent planet pairs are 16.7, 15.4, and 15.3 for the four-, five- and six-planet systems, respectively.} 
\label{mass-ecc}
\end{center} 
\end{figure*} 



\bsp	
\label{lastpage}
\end{document}